\documentclass[twocolumn,aps,prl,showpacs,preprintnumbers]{revtex4}%
\usepackage{mathrsfs}
\usepackage{amsmath}
\usepackage{amsfonts}
\usepackage{amssymb}
\usepackage{amsthm}
\usepackage{graphicx}
\usepackage{color}%
\providecommand{\U}[1]{\protect\rule{.1in}{.1in}}
\begin{document}
\title{Domain wall heat conductance in ferromagnetic wires}
\author{Peng Yan$^{1}$}
\author{Gerrit E.W. Bauer$^{2,1}$}
\affiliation{$^{1}$Kavli Institute of NanoScience, Delft University of Technology,
Lorentzweg 1, 2628 CJ Delft, The Netherlands}
\affiliation{$^{2}$Institute for Materials Research, Tohoku University, Sendai 980-8577, Japan}

\begin{abstract}
We present a theoretical study of heat transport in electrically insulating
ferromagnetic wires containing a domain wall. In the regime of validity of
continuum micromagnetism a domain wall is found to have no effect on the heat
conductance. However, spin waves are found to be reflected by domain walls
with widths of a few lattice spacings, which is associated with emergence of
an additional spin wave bound state. The resulting domain wall heat
conductance should be significant for thin films of Yttrium Iron Garnet with
sharply defined magnetic domains.

\end{abstract}

\maketitle

%75.60.Jk Magnetization reversal mechanisms\\
%75.75.+a Magnetic properties of nanostructures \\
%75.60.Ch Domain walls and domain structure
%85.75.-d Magnetoelectronics; spintronics:
%devices exploiting spin polarized transport or integrated magnetic fields
%85.70.Ay Magnetic device characterization, design and modeling
%75.30.Ds Spin waves\\
%----------------------------------------------------------------%
Spin wave (SW) excitations in magnetic systems have been studied for many
decades, but due to recent progress in the fabrication of magnetic
nanostructures, novel detection techniques, and new discoveries such as
current-induced magnetization dynamics, the field is very much alive
\cite{Demokritov}. It has been shown that SW can deliver signal information
and probe magnetic properties. Several device concepts have been proposed to
use SW logic devices \cite{Schneider,Khitun}. Recent experiments show that
three crucial material parameters, namely, the spin polarization $P$, the
intrinsic Gilbert damping $\alpha$, and the dissipative correction to
adiabaticity $\beta$, can be simultaneously determined by measuring the
current-induced SW Doppler effect \cite{Vlaminck,Koji}. Non-volatile data
storage devices \cite{Parkin} and logic circuits \cite{Cowburn} make use of
magnetic domain walls (DWs). The combination of both strategies is promising
as well \cite{Hertel}. Several phenomenon based on the interaction between SWs
and DWs have been proposed within the framework of continuum micromagnetic
theory, such absence of reflection \cite{Bayer} but an associated scattering
phases shift \cite{Hertel,Bayer} of SWs by a DW, frequency doubling
\cite{Sebastian}, domain wall drift \cite{Mikhailov,Han,Jamali}, and magnonic
spin transfer torque \cite{Yan,Hinzke,Kovalev}. A complication that has
received much less attention is the breakdown of the continuum approximation
of magnetization in materials with high anisotropy, in which the DW width
$\delta$ can be as small as a few lattice constants $a$ \cite{Jenkins}.
Atomic-scale DWs can display very different static and dynamic properties than
predicted by a continuum model \cite{Hilzinger}. For example, discrete DW
jumps that match the lattice periodicity have been observed by Novoselov
\emph{et al.} in thin films of yttrium-iron garnet (YIG) that support DWs as
narrow as $\delta/a\approx6/\pi$ \cite{Novoselov}. Only one attempt to
formulate the theory of SWs propagating through ultranarrow DWs is known to us
\cite{Liu}, predicting that a ferromagnetic SW can pass through a DW with
little reflection if its wavelength is less than twice the thickness of the
wall \cite{Liu}, which does not agree with our findings reported here.

\begin{figure}[ptbh]
\begin{center}
\includegraphics[width=8.5cm]{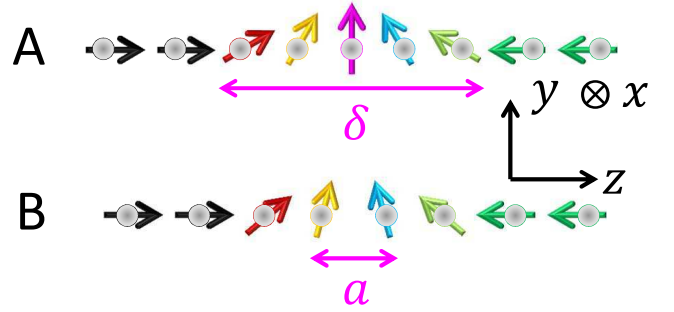}
\end{center}
\caption{(Color online) Principle ground state spin configurations for a
symmetric DW: its center either coincides with one of the lattice sites (type
A) or lies between them (type B). $\delta$ is the domain wall width and $a$
the lattice constant.}%
\end{figure}

In this Letter, we address DWs in wires of insulating ferromagnets such a YIG
that have superior magnetic quality factors, with damping constants that can
be three orders of magnitude smaller than those in metallic magnets. Their
application to novel memory concepts therefore appears attractive. Charge
current-induced DW motion is ruled out, of course, but heat currents can serve
identical purposes by virtue of the thermal spin-transfer torque
\cite{Hatami,Yan}. Here we study SW excitations in ferromagnetically coupled
spin chains in the presence of a DW, predicting remarkable effects of the
discrete lattice on the SW transmission. While we confirm previous results
that in a continuum model for DWs the SW transmission probability is unity,
spin wave reflection in a discrete spin model is found to be finite for abrupt
domain walls. The resulting domain wall heat conductance should help to detect
and manipulate DWs in insulating ferromagnetic wires.

We consider a wire that is sufficiently thin such that the lateral degrees of
freedom are frozen out and a 1D model is appropriate. Our model is a classical
Heisenberg spin chain with local anisotropy and nearest-neighbor ferromagnetic
exchange interaction, which is appropriate for materials such as YIG which has
a large effective spin $(S\approx14.3)$ per unit cell \cite{Tupitsyn}. In the
Hamiltonian \cite{Wieser1}
\begin{equation}
\mathcal{H}=-J\sum_{\left\langle n,m\right\rangle }\vec{S}_{n}\cdot\vec{S}%
_{m}-D\sum_{n}\left(  S_{n}^{z}\right)  ^{2},\label{Hamiltonian}%
\end{equation}
the first term describes the nearest neighbor interaction with ferromagnetic
coupling $J>0$, while the second terms is a local easy-axis anisotropy in the
$z$-direction with $D>0$. The local spin variable $\vec{S}_{n}$ is a
three-component unit vector on the $n$-th lattice site. A transverse hard axis
(shape) anisotropy causes complicated spin wave dispersions in thin films. In
our one-dimensional system, the dipolar interaction can be important as well,
but only leads to an increased easy-axis anisotropy \cite{Wieser1,Wieser2}.
For materials with a large unit cell like YIG, our approximation to integrate
the magnetic moment distribution in a single unit cell and replace them with
one classical spin is a good approximation at not too high temperatures
\cite{Cherepanov}. In insulating ferromagnets, the heat current is associated
with both phonon and magnon channels. In YIG the magnons, responsible for
interactions with DW of interest here, contribute $2/3$ of the total heat
current at low temperatures \cite{Rives, Douglass}. We consider a magnetic
wire of length $L$ that connects two large heat reservoirs held at two
temperatures $T_{L},T_{R},$ with constant difference $\Delta T=\left(
T_{L}-T_{R}\right)  >0.$ The magnon-mediated heat current carried by the spin
chain is given by the Landauer-B\"{u}ttiker formula \cite{Loss}
\begin{align}
\dot{Q} &  =\frac{1}{L}\int dk\rho_{k}\upsilon_{k}\hslash\omega_{k}\left\vert
t_{k}\right\vert ^{2}\left[  n_{B}\left(  \omega_{k},T_{L}\right)
-n_{B}\left(  \omega_{k},T_{R}\right)  \right]  \nonumber\\
&  =\frac{1}{2\pi}\int d\omega\hslash\omega\left\vert t_{k\left(
\omega\right)  }\right\vert ^{2}\left[  n_{B}\left(  \omega,T_{L}\right)
-n_{B}\left(  \omega,T_{R}\right)  \right]  .
\end{align}
Here $\rho_{k}=L/2\pi$ is the magnon density of states, $\upsilon_{k}%
=\partial\omega/\partial k$ is the SW group velocity, $\hslash\omega
_{k}=2D+2J\left[  1-\cos\left(  ka\right)  \right]  $ is the SW spectrum,
$t_{k}$ is the SW transmission coefficient, and $n_{B}\left(  \omega
,T_{L,R}\right)  =1/\left(  e^{\hslash\omega/\left(  k_{B}T_{L,R}\right)
}-1\right)  $ is the Bose-Einstein distribution. In the absence of DWs,
$t_{k}=1$ and for a small temperature bias%
\begin{equation}
\dot{Q}_{1}\dot{=}\frac{\Delta T}{2\pi}\int d\omega\hslash\omega\frac{\partial
n_{B}\left(  \omega,T\right)  }{\partial T},
\end{equation}
where $T=\left(  T_{L}+T_{R}\right)  /2.$

The heat current in the presence of DW is determined by the wall profile which
is obtained by energy minimization with respect to the polar angle $\theta
_{n}$ of $\vec{S}_{n}$ with the easy axis:%
\begin{equation}
\sin\left(  \theta_{n}-\theta_{n-1}\right)  -\sin\left(  \theta_{n+1}%
-\theta_{n}\right)  +\frac{D}{J}\sin\left(  2\theta_{n}\right)  =0,
\label{Equation}%
\end{equation}
where we chose that boundary conditions that spins are oriented in the $\pm z$
directions at the ends of the wire. The ground state DW configuration can be
computed iteratively \cite{Fullerton}. To second order in the small parameter
$D/J\ll1$ this procedure yields the Walker solution $\ln\tan\left(
\theta/2\right)  =z/\delta$ with DW width $\delta=a\sqrt{J/\left(  2D\right)
}$ \cite{Walker}. In a typical Walker DW, labeled A (metastable) in Fig. 1,
the spin of the central lattice site is oriented normal to the easy axis.
However, numerical calculation leads to a stable type B wall (Fig. 1) which
has always lower energies \cite{Hilzinger}. Clearly, this difference is
immaterial in the continuum limit for the magnetization (micromagnetics), and
it is only relevant for atomically sharp domain walls. We will discuss below
that the atomic-scale magnetization texture is observable in the heat and spin
transport properties.

Proceeding from the ground state DW configuration we can construct, linearize,
and solve the equations of motion $d\vec{S}_{n}/dt=-\vec{S}_{n}\times\left(
-\delta\mathcal{H}/\delta\vec{S}_{n}\right)  $ for the spins at a site $n$ in
order to determine the frequencies and amplitudes of the allowed SW modes.
This can be done conveniently in a coordinate system rotated about the
$x$-axis such that the transformed spins at equilibrium point to the new
$Z$-axis \cite{Tatara}, where the spin vectors in the two coordinate systems
are related as%
\begin{equation}
\left(
\begin{array}
[c]{c}%
S_{n}^{x}\\
S_{n}^{y}\\
S_{n}^{z}%
\end{array}
\right)  =\left(
\begin{array}
[c]{ccc}%
1 & 0 & 0\\
0 & \cos\theta_{n} & \sin\theta_{n}\\
0 & -\sin\theta_{n} & \cos\theta_{n}%
\end{array}
\right)  \left(
\begin{array}
[c]{c}%
S_{n}^{X}\\
S_{n}^{Y}\\
S_{n}^{Z}%
\end{array}
\right)  .
\end{equation}
The low-energy excitations correspond now to a small-angle precession around
the $Z$-axis with $S_{n}^{Z}\sim1$ \cite{Kovalev,Dugaev}$.$ The equations of
motion for the SW amplitude at site $n$ and frequency $\omega$ in the rotated
frame read%
\begin{align}
-i\frac{\omega}{J}S_{n}^{X}  &  =t_{n-1}S_{n-1}^{Y}+t_{n}S_{n+1}%
^{Y}\nonumber\\
&  -\left[  t_{n-1}+t_{n}-\frac{2D}{J}\left(  \sin^{2}\theta_{n}-\cos
^{2}\theta_{n}\right)  \right]  S_{n}^{Y},\nonumber\\
-i\frac{\omega}{J}S_{n}^{Y}  &  =-S_{n-1}^{X}-S_{n+1}^{X}\nonumber\\
&  +\left(  t_{n-1}+t_{n}+\frac{2D}{J}\cos^{2}\theta_{n}\right)  S_{n}^{X},
\end{align}
where $t_{n}=\cos\left(  \theta_{n}-\theta_{n+1}\right)  $. We obtain the
eigenfrequencies $\omega$ and amplitudes $S_{n}^{X},S_{n}^{Y}$ numerically.
\begin{figure}[ptbh]
\begin{center}
\includegraphics[width=8.5cm]{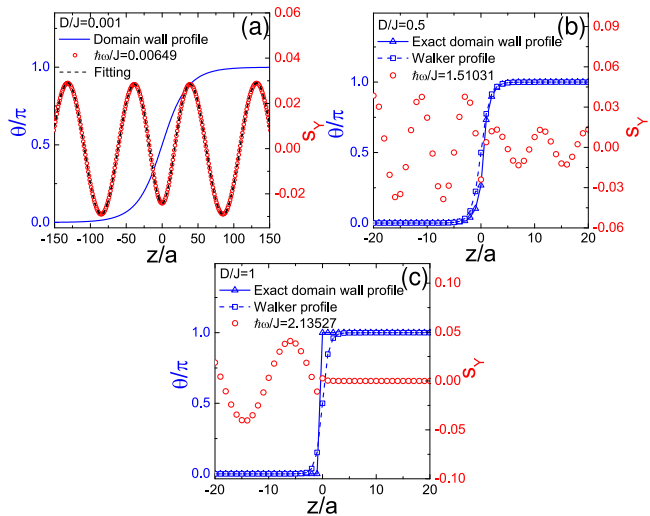}
\end{center}
\caption{(Color online) (a) The magnetization profile of a wide head-to-head
DW (solid line). The numerical SW solution (circles) for frequency $\omega$ in
a linear ferromagnetic chain cannot be distinguished from the solution (dashed
line) for the continuum model. (b) A computed narrow DW profile (triangles)
compared with the Walker model (squares). The SW amplitude for a selected
frequency is shown by circles. (c) The ground state profile of an abrupt DW
(triangles) compared with the Walker profile (squares). The amplitude of a SW
for another given frequency is shown by circles.}%
\end{figure}

$D/J=0.001$ leads to a broad DW profile (solid line in Fig. 2(a)) that cannot
be distinguished from the Walker solution. The SW mode profile $S_{n}^{Y}$ is
also plotted (circles), with the corresponding frequency given in the plot.
For a broad DW the SW profile agrees very well with the analytical solution
(dashed line in Fig. 2(a)) of the continuum model \cite{Yan} governed by the
Schr\"{o}dinger equation\textit{ }
\begin{equation}
\left[  -\frac{d^{2}}{d\xi^{2}}+V\left(  \xi\right)  \right]  \varphi\left(
\xi\right)  =q^{2}\varphi\left(  \xi\right)  , \label{Schrodinger}%
\end{equation}
with $\xi=z/\delta,$ $\varphi=S_{X}-iS_{Y},$ $q=k\delta$ with spin-wave vector
$k,$ and potential $V\left(  \xi\right)  =-2\sec$h$^{2}\xi.$ The potential
obtained here originates from the non-commutativity of the gauge
transformation with the momentum operator \cite{Kovalev}, similar to electron
transport in magnetic textures \cite{Brataas}. However, in contrast to
electrons, SWs do not feel a vector potential and only a scalar potential
remains. The solutions of Eq. (\ref{Schrodinger}) are one bound SW state and
propagating SWs that are not reflected at the DW but pick up a wave vector
$k$-dependent phase shift $\eta\left(  k\right)  =2\tan^{-1}\left\{  1/\left(
k\delta\right)  \right\}  $ \cite{Yan} that modifies the frequency dispersion
\cite{Wieser3} and causes interference effects in rings \cite{Hertel}. Since
the DW\ transmission coefficient is $t_{k}=e^{i\eta\left(  k\right)  }$ and
the transmission probability $\left\vert t_{k}\right\vert ^{2}=1,$ there is no
effect on DC heat transport. This conclusion is not modified by a transverse
hard-axis anisotropy $D_{\perp}\sum_{n}\left(  S_{n}^{x}\right)  ^{2}$, since
propagating SWs are elliptical but still reflectionless. The conclusion holds
in spite of the bound magnon state at the DW and in contrast to conducting
ferromagnets, where the DW leads to an increase of electric resistivity even
in the continuum model \cite{Ferrer}. In order to detect such a DW bound state
in insulating ferromagnets, one has to resort to spectroscopic or other
techniques \cite{Koma}.

The continuum approach breaks down for atomic scale DWs such as in Fig. 1. The
profiles of a relatively narrow DW for both exact numerical (triangles) and
Walker solution (squares) are shown in Fig. 2(b) for a large anisotropy
$D/J=0.5$. We see that now spin waves do get reflected! Figure 2(c) shows the
configuration of an abrupt DW (triangles), with $D/J=1.$ The Walker profile is
also plotted for comparison (squares). Interestingly, we now find that all SWs
(circles) are totally reflected. The physics is explained below.
\begin{figure}[ptbh]
\begin{center}
\includegraphics[width=8.5cm]{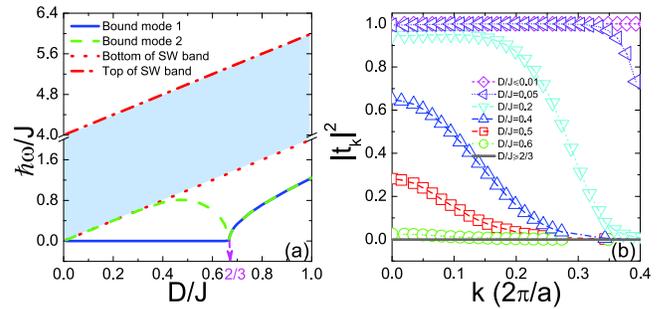}
\end{center}
\caption{(Color online) (a) Frequencies of DW bound states and SW continuum
(shaded area between dotted and dash-dotted lines). The solid curve represents
the lowest mode, while the dashed curve shows the second bound state. (b)
$k$-dependence of SW transmission for different values of $D/J$.}%
\end{figure}

As pointed out before, the DW supports bound SW states. Figure 3(a) shows the
numerical results for the bound state frequencies as a function of $D/J$ below
the continuum SW region, where the dotted line delineates the SW gap $\left(
\hslash\omega_{g}=2D\right)  $ and the dash-dotted line the top of the SW band
$\left(  \hslash\omega_{t}=2D+4J\right)  $. We confirm the existence of a
transition point at $D/J=2/3\ $that separates smooth and sharp DW
configurations \cite{Barbara} but without assuming $\left\vert \theta_{n\pm
1}-\theta_{n}\right\vert \ll1$, which is clearly not the case for these narrow
DWs. The transition point obtained here differs from the value $D/J=5/9\ $%
calculated in Ref. \cite{Vindigni}. Bound state $1$ (solid curve), below the
transition point $\omega_{b1}=0$ is caused by the rotational symmetry of the
smooth DW about the $z$ axis. When $D/J>2/3$ the DW width is the lattice
constant and bound state $1$ becomes degenerate with bound state $2$ (dashed
curve),\textit{ }i.e., $\omega_{b1}=\omega_{b2},$ with frequency monotonously
increasing with $D/J.$ For wide DWs, i.e., $D/J\ll2/3,$ bound state $2$ merges
with the continuum states. As the DW becomes narrower (increasing $D/J$),
$\omega_{b2}$ splits from the SW continuum and displays a non-monotonic
dependence on $D/J,$ with maximum at $\hslash\omega_{b2}\approx0.47D$,
dropping to zero when $D/J\rightarrow2/3$.

We can now understand why the SW is reflected at ultra-narrow DWs. Let us
consider a confining potential $V\left(  \xi\right)  \rightarrow0$ as
$\left\vert \xi\right\vert \rightarrow\infty$ for which the eigenvalue problem
Eq. (\ref{Schrodinger}) has $N$ bound states with $q=i\kappa_{n},$
$n=1,2,...,N,$ where $\kappa_{n}$ is real. Through the inverse scattering
formalism, one can construct a general $2N$-parameter formula for a
\emph{reflectionless} potential with $N$ bound states \cite{Thacker},%
\begin{equation}
V\left(  \xi\right)  =-2\frac{d^{2}}{d\xi^{2}}\ln\det A\left(  \xi\right)  ,
\label{Potential}%
\end{equation}
where the matrix $A_{mn}\left(  \xi\right)  =\delta_{mn}+\lambda_{m}%
\lambda_{n}/\left(  \kappa_{m}+\kappa_{n}\right)  ,$ with Kronecker function
$\delta_{mn}$ and $\lambda_{n}\left(  \xi\right)  =c_{n}e^{-\kappa_{n}\xi}$ is
symmetric$.$ For $N=1$ we recover the previous result of the continuum model
\cite{Yan}. In the discrete model, however, $N=2.$ Figure 3(a) clearly shows
that the bound state $2$ vanishes into the SW continuum only for wide DWs
($D/J\ll2/3$). Only in this limit there is just one bound state (mode $1$) and
the DW profile is perfectly described by the reflectionless Walker model. To
model the profile of a narrow DW, we expand Eq. (\ref{Equation}) to the fourth
order, and obtain the DW configuration $\ln\tan\left(  \theta/2\right)
=\left(  z/\delta\right)  +\left(  1/24\right)  \left(  a/\delta\right)
^{2}\left[  6\tanh\left(  z/\delta\right)  -z/\delta\right]  $ leading to the
confining potential $\tilde{V}\left(  \xi\right)  =-2\sec$h$^{2}\left[
y\left(  \xi\right)  \right]  $ with $y\left(  \xi\right)  =\xi+\left(
1/24\right)  \left(  a/\delta\right)  ^{2}\left(  6\tanh\xi-\xi\right)  .$
According to Thacker \emph{et al.} \cite{Thacker}, a reflectionless potential
with two bound states \emph{must} obey Eq. (\ref{Potential}). However, the
confining potential induced by a narrow DW as derived above cannot be
described by a four-parameter reflectionless potential Eq. (\ref{Potential}),
and indeed necessarily induces reflections. An expansion up to higher order
will not change this conclusion. Therefore, the SW must lose its property of
total transmission, as confirmed by our numerical results.

The $k$-dependent SW transmission for different parameters $D/J$ are shown in
Fig. 3(b), where the $k$ values are obtained by fast Fourier transformation of
the computed SW mode amplitudes \cite{Wieser2}. The wave vectors $k$ of
interest range here from zero up to $0.8\pi/a$ corresponding to a very short
wavelength ($2.5a$) beyond which a classical spin model might become
unreliable \cite{Prokop}.\textbf{ }In agreement with previous results there is
no detectable reflection in the calculations with $D/J<0.01.$ However, for a
large anisotropy, the SW will be only partially transmitted through the DW,
with transmission probability monotonically decreasing with increasing wave
vector $k.$ These results are very different from those obtained with a local
spin-spiral approximation, \textit{i.e.}, a constant pitch within the DW width
\cite{Gorkom}, which leads to perfect transmission for SWs with large $k$, but
hindered propagation at long wave lengths \cite{Liu}, which is caused by the
unphysical kinks in the magnetization texture of that model. We also have a
message for micromagnetic simulations in which the continuum Hamiltonian
$\mathcal{H}=J^{\prime}\left(  \partial\mathbf{m}/\partial z\right)
^{2}/2-Dm_{z}^{2}\ $is discretized into block spins $J^{\prime}\partial
^{2}\mathbf{m}_{n}/\partial z^{2}=J^{\prime}\left(  \mathbf{m}_{n-1}%
\mathbf{-}2\mathbf{m}_{n}\mathbf{+m}_{n+1}\right)  /\left(  \Delta z\right)
^{2}.$ The mesh size $\Delta z$ is found here to be rather critical, since SW
reflections that are artifacts in a continuum model occur unless $\pi
\sqrt{J^{\prime}/2D}\geqslant22$ $\Delta z.$

Our results should be relevant for systems with close to atomically sharp DW
such as YIG thin films. In the case of a narrow atomic-scale DW, the spin wave
gets reflections which leads to a modification to the heat transport as
$\dot{Q}_{2}=\dot{Q}_{1}+\Delta\dot{Q},$ where
\begin{equation}
\Delta\dot{Q}=\frac{\Delta T}{2\pi}\int d\omega\left(  \left\vert t_{\omega
}\right\vert ^{2}-1\right)  \hslash\omega\frac{\partial n_{B}\left(
\omega,T\right)  }{\partial T}.
\end{equation}
We can see that the DW decreases the heat conductance since $\Delta\dot{Q}<0$.
We can estimate the reduction of the heat current due to magnon reflection at
low temperatures as follows: magnons then exist only the bottom of SW band
($k=0$) at which $\left\vert t\left(  k=0\right)  \right\vert ^{2}$ can be
used to estimate the total transmission probability and $\Delta\dot{Q}/\dot
{Q}_{1}\approx\left\vert t\left(  k=0\right)  \right\vert ^{2}-1.$ When
$D/J=0.4$, $\Delta\dot{Q}/\dot{Q}_{1}\sim-35\%$. At room temperature $k_{B}T$
is larger than the exchange energy $J\approx40$ K in YIG
\cite{Tupitsyn,Cherepanov} and magnons with high wave numbers become important
for the heat transport. The spin wave reflection becomes significant when
$\delta\lesssim\left(  10/\pi\right)  a,$ where $a=1.24$ nm in YIG
\cite{Novoselov,Tupitsyn,Cherepanov}.

Ultranarrow DWs are sensitive to crystal lattice pinning: the A and B-type
walls in the 1D model (Fig. 1) have different ground state energies
\cite{Hilzinger}. A domain wall therefore has to surmount an energy barrier
when moving from one atomic plane to another in a \textquotedblleft
Peierls\textquotedblright\ potential and a threshold external force is
necessary to assist the DW motion \cite{Hilzinger}. Novoselov \emph{et al}.
\cite{Novoselov} indeed observed a critical magnetic field for field-driven
atomic-scale DW dynamics. Consequently, an intrinsic critical heat/spin
current exists for the propagation of DWs even in perfect wires. According to
Ref. \cite{Hinzke}, a temperature gradient of $1$ K/$%
%TCIMACRO{\unit{nm}}%
%BeginExpansion
\operatorname{nm}%
%EndExpansion
$ creates a pressure corresponding to a field $H=5$ $%
%TCIMACRO{\unit{mT}}%
%BeginExpansion
\operatorname{mT}%
%EndExpansion
$ for a magnetic insulator with a saturation magnetization $M_{s}%
=2\times10^{6}$ A/m. We estimate the critical temperature gradient to overcome
the atomic pinning as $0.1$ K/$%
%TCIMACRO{\unit{\U{3bc}m}}%
%BeginExpansion
\operatorname{\mu m}%
%EndExpansion
$ for the critical magnetic field value $H_{c}=0.7$ G and $M_{s}%
=1.6\times10^{4}$ A/m reported in Ref. \cite{Novoselov}.

In conclusion, we studied the heat transport by SWs through magnetic DWs. When
the DW becomes the order of a few lattice constants a second bound state
emerges that is absent in the continuum approximation and causes strong SW
reflection. The SW reflection leads to a DW heat conductance. The results
developed here are relevant for ultrahigh-density storage devices based on DWs
\cite{NEC} when based on insulating ferromagnets.

We thank Akashdeep Kamra for helpful discussion. This work is supported by the
FOM foundation, DFG Priority Program 1538 SpinCat, and EG-STREP MACALO.

\end{document}